\documentclass[aps,twocolumn,prb]{revtex4}

\usepackage{graphicx}
\usepackage{amssymb}
\usepackage{amsmath}
\newcommand{\R}{\mathbf{r}}
\newcommand{\UP}{n_{\uparrow}}
\newcommand{\DN}{n_{\downarrow}}

\begin{document}

\title{Global hybrids from the semiclassical atom theory satisfying the local density linear response}
\author{Eduardo Fabiano}
\affiliation{National Nanotechnology Laboratory (NNL), Istituto Nanoscienze-CNR, Via per Arnesano 16, I-73100 Lecce, Italy }
\affiliation{Center for Biomolecular Nanotechnologies @UNILE, Istituto Italiano di Tecnologia, Via Barsanti, I-73010 Arnesano, Italy}
\email{eduardo.fabiano@nano.cnr.it}
\author{Lucian A. Constantin}
\affiliation{Center for Biomolecular Nanotechnologies @UNILE, Istituto Italiano di Tecnologia, Via Barsanti, I-73010 Arnesano, Italy}
\author{Pietro Cortona}
\affiliation{Laboratoire Structures, Propri\'et\'es et Mod\'elisation des Solides, CNRS UMR 8580, \'Ecole Centrale Paris, Grand Voie des Vignes, F-92295 Ch\^{a}tenay-Malabry, France}
\author{Fabio Della Sala}
\affiliation{National Nanotechnology Laboratory (NNL), Istituto Nanoscienze-CNR, Via per Arnesano 16, I-73100 Lecce, Italy }
\affiliation{Center for Biomolecular Nanotechnologies @UNILE, Istituto Italiano di Tecnologia, Via Barsanti, I-73010 Arnesano, Italy}
\date{\today}

\begin{abstract}
We propose 
global hybrid approximations of the exchange-correlation (XC) energy
functional which
reproduce well the modified fourth-order gradient expansion of the exchange energy 
in the semiclassical limit of many-electron neutral atoms
and recover the 
full local density approximation (LDA) linear response.
These XC functionals represent the hybrid versions of the APBE functional 
[Phys. Rev. Lett. 106, 186406, (2011)] yet employing an 
additional correlation functional which uses the localization 
concept of the correlation energy density to improve the compatibility
with the Hartree-Fock exchange as well as  the coupling-constant-resolved XC potential energy. 
Broad energetical and structural testings, 
including thermochemistry and geometry, transition metal complexes, 
non-covalent interactions, gold clusters and small gold-molecule interfaces, 
as well as an analysis of the hybrid parameters, show that
our construction is quite robust. 
In particular, our testing shows that 
the resulting hybrid, including 20\% of Hartree-Fock exchange and named hAPBE,
performs remarkably well for a broad palette of systems and properties, being generally 
better than popular hybrids (PBE0 and B3LYP).
Semi-empirical dispersion corrections are also provided.
\end{abstract}

\maketitle

\section{Introduction}
Density functional theory (DFT) \cite{hohen,sham}
is nowadays one of the most popular methods in electronic structure calculations.
DFT is an exact theory, but its practical implementation requires 
an approximation for the so-called exchange-correlation (XC) energy
functional, which describes the quantum effects of the 
electron-electron interaction.  
Many approximated expressions, having different levels of 
complexity/sophistication, exist for this functional \cite{ladder,scuseria_rev}. 
They can be roughly classified in two broad classes: 
i) local and semilocal functionals, using as input information the
electron density, its derivatives, and the kinetic energy density; 
ii) fully non-local functionals,
which additionally use as input non-local quantities computed from the Kohn-Sham
orbitals (e.g. exact exchange). 
The former ones, which include local density, 
generalized gradient and meta generalized gradient approximations 
(LDA, GGAs, and meta-GGAs, respectively), are fast and sufficiently accurate 
for many purposes \cite{TCA,RevTCA,TCA1,bloc,js,blochole,am05_1,am05_2,pittalis09,MdelCampo2012179,kcisk,m06-l,mggams,q2D,peri,inter}, 
but fail for a number of important properties such as 
density distributions or barriers heights 
\cite{janesko08,cohen12,hpbeint,grabowski11,grabowski14}. 
The second ones, which include 
global, range-separated, local and double hybrid functionals
\cite{becke1,becke2,savin,hirao,scuseria,m06,m05,pbe0,pbe0_2,pbe0_3,pbe13,guido13,hermet,hybrid1,hybrid2,hybrid3,hybrid4,hybrid5,hybrid6,
hybrid7,hyper1,hyper2,hyper3,hyper4,hyper5,hyper6,hyper7,hyper8,hyper9,cohen12,double1,double2,double3}, improve 
considerably the results, 
but they are much more computer-time demanding, in particular when 
local or double hybrids are used.

Among 
functionals of the second class, 
global hybrids provide the best compromise 
between accuracy and 
efficiency. They are characterized by an 
exchange energy contribution which is a mixing of exact exchange 
and a local or semilocal contribution. 
In global hybrids, a fraction of Hartree-Fock exchange energy is directly 
combined with the complementary fraction of an exchange energy functional. 

The great majority of the hybrid functionals are parameterized: 
the constant determining the relative weight of the exact and semilocal exchange 
and (in most cases) some other parameters are fixed by fitting some 
training datasets.  Furthermore, fitted parameters can also be contained 
in the semilocal functional used in the hybrid construction.  
A well-known example of parameterized hybrid functional is 
B3LYP \cite{b88,lyp,b3lyp,b3lyp_2}. 

On the other hand, there are few examples of non-parameterized
hybrid functionals. In the case of global hybrids 
(the one of interest in the present paper) the parameter determining 
the ratio between exact and semilocal exchange can be chosen 
by theoretical arguments \cite{pbe0,pbe13}. Then, constructing the 
hybrid on the basis of a non-parameterized semilocal functional, 
one obtains a parameter-free (or non-empirical) hybrid functional.  
PBE0 \cite{pbe0,pbe0_2,pbe0_3} and, more recently, PBE0-1/3 \cite{pbe13,guido13} 
have been obtained in such a way. 
Extensive calculations of molecular properties have shown 
that non-parameterized hybrids are not only more satisfactory 
than the parameterized ones from the theoretical point of view, but 
they are also competitive in terms of accuracy, 
in particular when systems or properties that do not belong 
to the training sets of the parameterized hybrids are considered.

In the present paper, we construct and test parameter-free hybrids based on 
the APBE GGA functional \cite{apbe}. This is a non-parameterized GGA constructed
from the semiclassical atom model and has shown high accuracy for a wide range
of molecular  properties \cite{mukappa}. Moreover, in our development of the
hybrid form, we deviate from the conventional scheme and 
explore the possibility of combining the exact exchange with 
an additional 
correlation functional. 
This procedure is motivated,
within the adiabatic connection scheme \cite{becke1,pbe0,pbe13}, by
the need to correct the correlation contribution associated to the
exact exchange fraction, when the coupling constant is approaching one.
A similar motivation stands at the base of double hybrids, 
where the correlation is however partly treated at the MP2 level.

The paper is organized as follows: in the next section
we present the theory at the base of our construction of the
hybrid functionals and discuss it in the context of the
adiabatic connection. A possible spin-dependent
correction for the correlation is discussed as well.
In the following section we show the performance
of our hybrid functionals in comparison with other
relevant ones, using a broad set of molecular tests.
The results are discussed also in terms of the role 
of the parameters defining the hybrids. Finally,
we provide a brief study on the use of 
a semiempirical dispersion correction in conjuction
with the functionals presented in this work.

\section{Theory}
\subsection{Construction of the hybrid functional} \label{sec:con}
According to the adiabatic connection the exchange-correlation (XC)
functional is given by the coupling-constant integration formula
\begin{equation} 
\label{adiab}
E_{xc} = \int_0^1W_{xc,\lambda}d\lambda\ ,
\end{equation}
where $W_{xc,\lambda}$ is an appropriate coupling-constant-resolved
XC potential energy.
One popular choice for $W_{xc,\lambda}$
to derive global hybrid functional is
the ansatz proposed by Perdew, Ernzerhof, and Burke
\cite{pbe0}: 
\begin{equation} \label{PEBmod}
W_{xc,\lambda} = W_{xc,\lambda}^{DFA}+\left( E_x^{HF}-E_x^{DFA} \right) \left(1-\lambda \right)^{n-1} ,
\end{equation}
where DFA stands for some local or semilocal density functional approximation
and $E_x^{HF}$ is the Hartree-Fock exchange computed with Kohn-Sham
orbitals.
This formula correctly reverts to exact-exchange-only 
when $\lambda=0$ (i.e. for the non-interacting system) and assumes that $W_{xc,1}$ is well approximated by DFA\cite{pbe0,kaupp08}.
We recall \cite{pbe0} that $W_{xc,\lambda}^{DFA}=E_x^{DFA}+U_c^{DFA}(\lambda)$,
where
$U_c^{DFA}(\lambda)=\frac{d}{d\lambda} \{ \lambda^2 E_c^{DFA}[\rho_{\uparrow,\lambda^{-1}},\rho_{\downarrow,\lambda^{-1}}] \}$, with 
$\rho_{\sigma,\gamma}(\R)=\gamma^3 \rho_\sigma(\gamma\R)$ being the uniform scaled spin-densities.
Thus, $U_c^{DFA}(\lambda)$
is a density functional approximation
for the correlation potential 
($U_c^{DFA}(0)=0$ and 
$U_c^{DFA}(1)=E_c^{DFA}-T_c^{DFA}$), and Eq. (\ref{PEBmod})
describes, at small non-vanishing values of the coupling constant $\lambda$,
a mixing of the DFT exchange with the exact exchange, but with
full DFT correlation as given by DFA.
Using  Eq. (\ref{PEBmod}) in Eq. (\ref{adiab}) we have:
\begin{equation} 
\label{orig}
E_{xc} = \frac{1}{n} E_x^{HF}+\left(1-\frac{1}{n}\right) E_x^{DFA} +  E_c^{DFA}
\end{equation}
which shows that always the  same $E_c^{DFA}$ is used in combination with different value of $n$.
This situation  appears not to be optimal, since in general the DFA correlation contribution is designed to work well
together with its DFA exchange counterpart but not with the HF exchange.
Thus, it may be appropriate to generalize Eq. (\ref{orig}). 
Hence, we propose the generalized ansatz
\begin{eqnarray}\label{PEBmodg2}
W_{xc,\lambda}  & = & W_{xc,\lambda}^{DFA1}+\left(E_x^{HF} -E_{x}^{DFA1} \right) \left(1-\lambda 
\right)^{n-1}  \nonumber \\
& + & \left( E_c^{DFA2} -E_{c}^{DFA1} \right) \lambda^{m-1} ,
\end{eqnarray}
 so that after  $\lambda$ integration  we obtain
\begin{eqnarray}\label{Final1}
E_{xc}  & = & 
\frac{1}{n} E_x^{HF}+\left(1-\frac{1}{n}\right) E_x^{DFA1} + \\
\nonumber
&& + \frac{1}{m} E_c^{DFA2} + \left(1-\frac{1}{m}\right) E_c^{DFA1}  \, \, . 
\end{eqnarray}
where the integer parameter $m$ controls how the correlation  functionals $E_c^{DFA1}$  and $E_c^{DFA2}$ are mixed. 
We note that Eq. (\ref{PEBmodg2}) is exact at $\lambda=0$, as Eq.  (\ref{PEBmod}), while at $\lambda=1$ we have
  $W_{xc,1}= W_{xc,1}^{DFA1}+E_c^{DFA2}-E_c^{DFA1}$, differently from Eq. (\ref{PEBmod}).
 As shown in the Appendix, Eq. (\ref{PEBmodg2}) can yield  an improved coupling-constant-resolved XC potential for atoms.
In this work, we will choose for DFA1 the APBE GGA functional 
\cite{apbe}, which is one of the most accurate 
non-empirical GGAs with broad applicability
\cite{apbe,mukappa}.
For DFA2, we select the PBEloc correlation
functional \cite{loc}, which has been proved to work
quite well together with the Hartree-Fock exchange, even if 
other possibilities could also be considered \cite{zv,gaploc,RevTCA}.
Note that Eq. (\ref{Final1}) resembles 
the general expression of two-parameter double hybrid functionals \cite{savin11}, if $E_c^{DFA2}$ is replaced by the   
MP2 correlation energy and the constraint on $n$ and
$m$ to be integer numbers is relaxed. 

In order to use only one parameter, we need a relation between $n$ and $m$. 
We note that the APBE functional was constructed in such 
a way that it recovers the accurate LDA 
linear response behavior \cite{moroni95,ortiz92}.
Similarly the PBEloc functional was designed to respect
the same property with respect to exact exchange.
Thus, it appears a natural choice to force the full hybrid functional
to respect the accurate LDA 
linear response behavior. To this end we need to put
$m=n$ in Eq. (\ref{Final1}). The final formula for the
hybrid XC functional is therefore
\begin{equation}\label{e4}
E_{xc} = \frac{1}{n}\left(E_x^{HF}+E_c^{PBEloc}\right) + \left(1-\frac{1}{n}\right)E_{xc}^{APBE}\;\; .
\end{equation}
Equation (\ref{e4}) shows that the resulting XC energy is just a linear mixing 
of the original APBE XC functional with the $E_x^{HF}+E_c^{PBEloc}$ functional:
thus a good accuracy can be expected only if PBEloc works well in combination with HF, as it is indeed the case\cite{loc}.

We recall that the relevance of the LDA linear response
behavior has been already discussed
for the semilocal level of the theory \cite{pbe,apbe,mukappa,cancio12}, 
and even for hybrid functionals \cite{delcampo12}.
However, in Ref. \citenum{delcampo12} only
the linear response behavior of the semilocal fraction of
the functional is considered. 
We will discuss this item even further in the 
final part of this paper (Section \ref{sec:par}), where the condition $n=m$
will be released.

Moreover, we recall that $E_{x}^{APBE}$ was constructed 
from the semiclassical theory of large
neutral atoms, and recovers the modified 
second-order gradient expansion (MGE2) \cite{apbe}. 
It was proved that MGE2 can reproduce well 
the fourth-order gradient expansion (GE4) 
\cite{apbe,EB09,LCPB09}. Thus, because $E_x^{HF}$ already
recovers the GE4, the whole exchange part of the hybrid 
given in Eq. (\ref{e4}) reproduces to a good accuracy 
the GE4 in the slowly-varying atomic density limit. 
This feature however, can be met exactly only 
at the meta-GGA hybrid level (e.g. the TPSSh hybrid \cite{tpssh}).

The value of the remaining parameter $n$ can be strongly circumscribed by theoretical
considerations based on perturbation theory \cite{pbe0,pbe13} 
but not univocally fixed. The more convenient choice depends on the 
semilocal functional(s) used in the hybrid construction and is
ultimately determined by the result accuracy. In this work we consider $n=4$ or 5.
The corresponding functionals
are assessed in next sections.
These functionals have the same exact exchange content as the popular
PBE0 \cite{pbe0,pbe0_2,pbe0_3} and B3LYP \cite{b88,lyp,b3lyp,b3lyp_2}
hybrid functionals, respectively.
We note that additional values of $n$ can be considered. 
In particular, the value $n=3$, which can also be obtained from
perturbation theory \cite{pbe13}, was shown
to yield good results for organic chemistry \cite{guido13} 
and excited states, in particular when charge transfer \cite{guido2} and
Rydberg transitions \cite{guido3} are considered.
However, such higher fractions of exact exchange do not provide a global
result improvement for the datasets considered in the present paper, thus
they will not be discussed any longer in this work.

\subsection{Spin-dependent correction for the PBEloc functional}
The good compatibility of the PBEloc functional with the Hartree-Fock
exchange is based on a localization 
paradigm of the correlation energy density \cite{loc,bloc,blochole}. In Ref.
\citenum{zv} it has been shown that this localization, and consequently
the compatibility with exact exchange, can be further enhanced,
preserving the exact properties of PBEloc,
by a slight modification of the original correlation functional to
\begin{equation}
E_c^{zvPBEloc} = \int \rho(\R) f_{\alpha,\omega}(\zeta(\R),v(\R))\epsilon_c^{PBEloc}(\R)d\R\ ,
\end{equation}
where $\rho$ is the electron density,
$\epsilon_c^{PBEloc}$ is the PBEloc correlation energy 
per particle, and 
\begin{equation}\label{fzv}
f_{\alpha,\omega}(\zeta,v)= e^{-\alpha v^3|\zeta|^\omega},
\end{equation}
is a spin-dependent correction factor with
$v= |\nabla n|/(2k_v\rho)$ being a reduced gradient 
suitable for density variations in valence and
bonding regions \cite{zv}, $k_v=2 (3/(4\pi^4))^{1/18}\rho^{1/9}$, and 
$\zeta$ being the relative spin polarization.

%
\begin{figure}
\includegraphics[width=\columnwidth]{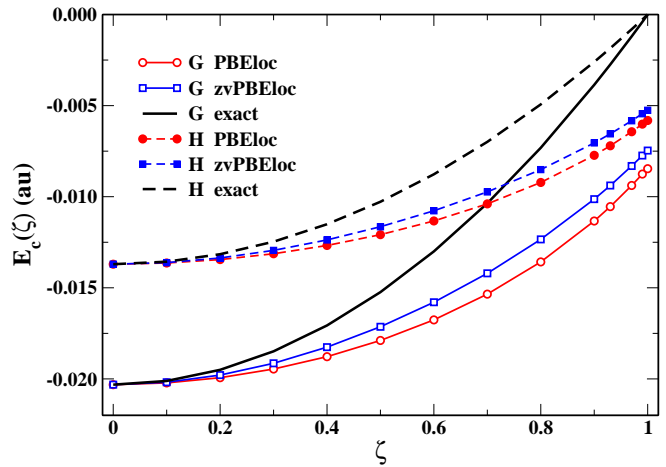}
\caption{ $E_c(\zeta)$ versus $\zeta$ of the one-electron Gaussian
and Hydrogenic densities with uniform spin-polarization $\zeta$,
for PBEloc, zvPBEloc, and ideal $E_c(z)$ of Eq. (\ref{e24}).}
\label{f5}  
\end{figure}
%
To fix the parameters $\alpha$ and $\omega$ in Eq. (\ref{fzv})
we used the uniformly spin-polarized Gaussian (G) and hydrogenic (H) one-electron
density models $\UP=\frac{1+\zeta}{2} n_{G,H}$ and $\DN=\frac{1-\zeta}{2}n_{G,H}$, with 
$n_H(r) = \frac{e^{-2r}}{\pi}$ and $n_G(r)=\frac{e^{-r^2}}{\pi^{3/2}}$ respectively, requiring 
an improved spin-behavior for $\zeta\ge 0.3$. We recall that the ideal spin-behavior for these 
model systems is \cite{zv}
\begin{equation}
E_c(\zeta)=E_c^{PBEloc}(\zeta =0)g(\zeta).
\label{e24}
\end{equation}
with $g(\zeta)=1-\zeta^2$. We finally obtain $\alpha=0.5$ and $\omega=2.0$.
As shown in Table \ref{tab1b} the so obtained correlation zvPBEloc
functional displays an improved compatibility with Hartree-Fock exchange,
with respect to the original PBEloc, for tests on small
molecules where dynamical correlation is relevant
(see Refs. \citenum{zv,zpbeint} for more details).
Moreover, Fig. \ref{f5} shows that 
zvPBEloc provides a better behavior with respect
to PBEloc, yet preserving the PBEloc shape, for both H and G densities.
%
\begin{table}[b]
\begin{center}
\caption{\label{tab1b} Mean absolute errors (kcal/mol) on 
AE6 test \cite{ae6} for atomization energies of organic molecules,
BH6 test \cite{k9} for barrier heights of organic molecules, and 
K9 test \cite{k9} for kinetics of organic molecules,  
as resulting from calculations
using Hartree-Fock exchange and different GGA correlation functionals.}
\begin{ruledtabular}
\begin{tabular}{lrrrr}
Test set & PBE & LYP & PBEloc & zvPBEloc\\
\hline
AE6 & 31.9    & 38.2 & 24.0 & 20.8   \\
BH6 & 5.6 & 5.3 & 4.4 & 4.6 \\
K9  & 5.7    & 6.0 & 4.7 & 4.2 \\
\end{tabular}
\end{ruledtabular}
\end{center}
\end{table}
%

To take advantage of the improved features of the zvPBEloc
functional we will thus consider also global hybrids of the form
\begin{equation}\label{e41}
E_{xc} = \frac{1}{n}\left(E_x^{HF}+E_c^{zvPBEloc}\right) + \left(1-\frac{1}{n}\right)E_{xc}^{APBE}\ .
\end{equation}

\section{Computational details}
To assess the performance of the functionals we performed calculations
on a large test set (overall more than 500 systems and various energetical and structural ground-state properties),
which was divided into five main groups:
\begin{itemize}
\item \textbf{Thermochemistry}. This group considers 
the atomization energies of small  molecules (AE6 \cite{ae6,ae62}
and W4 \cite{double1,grimme_test1,grimme_test2}) 
and molecules with non-single-reference character
(W4-MR \cite{double1}), proton affinities (PA12 
\cite{pa12_1,pa12_2,pa12_3,grimme_test1,grimme_test2}) and ionization
potentials (G21IP \cite{g21ip,grimme_test1,grimme_test2}), 
different barrier heights (BH76
\cite{bh76_1,bh76_2,grimme_test1,grimme_test2}), 
reaction energies (BH76RC 
\cite{bh76_1,bh76_2,grimme_test1,grimme_test2} and OMRE \cite{hpbeint}), 
and both barrier heights and reaction energies (K9 \cite{k9,ae62}).
\item \textbf{Organic molecules' geometry}. This group comprises
tests on bond lengths of hydrogenic (MGHBL9 \cite{mgbl19})
and non-hydrogenic (MGNHBL11 \cite{mgbl19}) bonds, bond lengths
of open-shell organic molecules (BL9 \cite{zv}) as well as
vibrational frequencies (F38 \cite{f38}) of small organic molecules. 
\item \textbf{Transition metals}. This group of tests includes
atomization energies of small transition metal complexes (TM10AE
\cite{3dmetals,zpbeint}) and gold clusters (AUnAE \cite{zpbeint,pbeint_gold},
reaction energies of transition metal complexes (TMRE \cite{hpbeint,3dmetals}),
and bond lengths of transition metal complexes (TMBL \cite{zpbeint,buhl06})
and gold clusters (AuBL6 \cite{hpbeint,pbeint_gold}).
Note that here for the TM10AE test have been used
tighter convergence criteria than in previous publications \cite{zpbeint,bloc}.
\item \textbf{Non-covalent interactions}. This group considers interaction
energies of hydrogen-bond (HB6 \cite{noncov}), dipole-dipole (DI6 \cite{noncov}), 
charge-transfer (CT7 \cite{noncov}), and $\pi$-$\pi$-stacking (pp5 \cite{noncov})
complexes. In addition, interaction energies of dihydrogen-bond complexes
(DHB23 \cite{dihydro}) and of complexes with various character (S22 \cite{s22,s222})
are taken into account.
\item \textbf{Other properties}. This group considers a miscellaneous of other properties 
and systems including isomerization energies of large organic molecules 
(ISOL6 \cite{isol6}), difficult cases for DFT (DC9/12 \cite{dc9}), small
gold-organic interfaces (SI12 \cite{hpbeint}), dipole moments of
organic molecules (DM25 \cite{hpbeint}), and atomic energies (AE17 \cite{m06}).
\end{itemize}

All tests were carried out for the functionals defined by Eqs. (\ref{e4})
and (\ref{e41}) with $n=$ 4 or 5.
For completeness also the hybrid functional named APBE0 and defined
as $E_{xc}^{APBE0} = 0.25E_x^{HF} + 0.75E_x^{APBE} + E_c^{APBE}$
was considered (this is in practice obtained from Eq. (\ref{e4}) changing
$E_c^{PBEloc}$ with $E_c^{APBE}$ and setting $n=4$
). 

For comparison we performed calculations using several other GGA and hybrid
functionals:  the PBE \cite{pbe} and APBE \cite{apbe} GGA functionals,
as well as the popular hybrid XC functionals
PBE0 \cite{pbe0,pbe0_2,pbe0_3} and B3LYP \cite{b88,lyp,b3lyp,b3lyp_2}.
Finally, we considered also the PBEmol$\beta$0 hybrid functional \cite{delcampo12}
which was constructed to restore the
LDA linear response in the semilocal part of the functional
by scaling the second-order coefficient $\beta$ in the correlation
part.
Note that, because the GGA functional PBEmol \cite{delcampo12}  
is very similar to APBE, the PBEmol$\beta$0 hybrid functional 
practically differs from Eq.  (\ref{e4}) only for the fact that  
in the former the LDA linear response condition is enforced only in the
semilocal part, while in the later it is extended to the
whole functionals (thanks to the inclusion of the PBEloc correction).

All calculations have been performed with the TURBOMOLE
program package \cite{turbomole} using a def2-TZVPP basis set
\cite{basis1,basis2}. The choice of the basis set was motivated by the need to find 
a best compromise between accuracy and computational cost, so that our results can 
be directly compared to practical applications. 
We note also that because the same basis set is used for all the functionals 
a fair assessment of relative performances is possible.

To evaluate the performance of the different approaches we computed,
for each group of tests outlined above, the
average MAE relative to PBE0, which is assumed as a reference.
Hence, we considered
\begin{equation}
\mathrm{RMAE} = \frac{1}{M}\sum_{i=1}^M\frac{\mathrm{MAE}_i}{\mathrm{MAE}_i^{PBE0}}
\end{equation}
where the sum runs over all the $M$ tests within a group and MAE$_i^{PBE0}$ 
is the MAE of PBE0 for the $i$-th test. The RMAE indicates
whether any method is better (RMAE$<$1) or worse (RMAE$>$1) than PBE0.
The RMAE provides a fair global assessment of all the results,
however it may tend to overweight tests where both MAE$_i$ and
MAE$_i^{PBE0}$ are small and to underweight the results of tests where
both methods yield quite poor results. For this reason, in addition
to the RMAE, we considered also the weighted MAE
\begin{eqnarray}
\label{e7}
\mathrm{WMAE} & = & \frac{1}{M}\sum_{i=1}^M\frac{\mathrm{MAE}_i-\mathrm{MAE}_i^{PBE0}}{\langle\mathrm{MAE}_i^{PBE0}\rangle} = \\
\nonumber
& = &  \frac{1}{M}\sum_{i=1}^M\left(\frac{\mathrm{MAE}_i}{\mathrm{MAE}_i^{PBE0}}-1\right)\frac{\mathrm{MAE}_i^{PBE0}}{\langle\mathrm{MAE}_i^{PBE0}\rangle}\ ,
\end{eqnarray}
where $\langle\mathrm{MAE}_i^{PBE0}\rangle$ is the 
average of the MAEs of all hybrid methods for test $i$.
Equation (\ref{e7}) shows that in the WMAE the 
quantities composing the RMAE are averaged with weights 
$w_i=\mathrm{MAE}_i^{PBE0}/\langle\mathrm{MAE}_i^{PBE0}\rangle$, which
indicate the relative performance of PBE0 (assumed as a reference) with respect to
what may be expected from hybrid approaches. Hence, negative WMAEs denote
methods outperforming PBE0, while positive values denote
methods performing worse than PBE0.

\section{Results}
\begin{table*}
\begin{center}
\caption{\label{tab1} Mean absolute errors (MAEs) for different tests and functionals. For each group of tests the average MAE relative to PBE0 and the weighted MAE (RMAE and WMAE; see text for definitions) are reported in the last lines. For each line the hybrid functional performing best is highlighted in bold face; a star is used to denote the best among the functionals based on Eqs. (\ref{e4}) and (\ref{e41}).}
\begin{ruledtabular}
\begin{small}
\begin{tabular}{lrrcrrrcrrrrr}
Test & \multicolumn{2}{c}{GGA} && \multicolumn{3}{c}{Hybrid $n=5$} && \multicolumn{5}{c}{Hybrid $n=4$} \\
\cline{2-3}\cline{5-7}\cline{9-13}
&   PBE  &  APBE  &&  B3LYP  &  Eq. (\ref{e4})  & Eq. (\ref{e41})   &&  PBE0  &  APBE0  &  PBEmol$\beta$0  &  Eq. (\ref{e4})  & Eq. (\ref{e41})  \\
\hline
\multicolumn{13}{c}{Thermochemistry (kcal/mol)} \\
atomization energy (AE6)  &  13.4  &  7.8  &&  5.5        & 5.8           & $^*${\bf 4.2}  && 6.3   &  9.5     &  7.3     &  6.8     & 4.8  \\
atomization energy (W4)   &  10.7  &  8.5  &&  {\bf 5.8}  &  6.4          &  $^*$6.0       &&  6.0  &  8.7     &  7.6     &  7.3     &  6.4  \\
static correlation (W4-MR)&  21.8  &  17.4 &&  8.4        & $^*${\bf 5.9} & $^*${\bf 5.9}  &&  7.1  &  10.5     &  11.2    &  9.1     &  7.2 \\
proton affinities (PA12)  &  2.2   &  2.8  &&  {\bf 2.3}  &  $^*$3.1       &  $^*$3.1      &&  2.8  &  3.4      &  3.5     &  3.3     &  3.3  \\
ionization pot. (G21IP)   &  3.9   &  4.0  &&  {\bf 3.8}  &  $^*$4.0       &  $^*$4.0      &&  4.1  &  4.2      &  4.6     &  4.2     &  4.2  \\
barrier heights (BH76)    &  9.8   &  9.0  &&  5.0        &  5.0           &  5.0          &&  4.5  &  {\bf 3.9}  & 4.0    &  $^*$4.2  &  $^*$4.2  \\
reaction energies (BH76RC)&  4.4   &  3.8  &&  2.6        &  2.8           &  2.9          &&  2.7   &  2.5      &  2.4    &  $^*$2.6  &  2.7  \\
reaction energies (OMRE)  &  6.7   &  7.1  && {\bf 4.6}   &  6.5           &  $^*$6.4      &&  9.1    &  7.1     &  6.0     &  8.1        &  7.9  \\
kinetics (K9)             &  7.5   &  6.6  &&  3.8        &  4.2           &  4.2           &&  3.9    &  3.3    &  {\bf 3.1}  &  $^*$3.7  &  $^*$3.7  \\
RMAE                      &  1.71  &  1.46 &&  {\bf 0.96} & 1.00           &  $^*${\bf 0.96} &&  1.00   &  1.15   &  1.09     &  1.07     &  0.99  \\
WMAE(\%)                  &  +66   &  +42  && {\bf -6}     &  -1           &  $^*$-4         &&  0     &  +11    &  +6        &  +6       & -2  \\
\multicolumn{13}{c}{Organic molecules' geometry (m\AA{}, and cm$^{-1}$)} \\
H bond lengths (MGHBL9)      &  11.4  &  10.2  &&  3.0       &  $^*$1.4        &  $^*$1.4          &&  2.3     &  1.5  &  {\bf 1.0}  &  1.5  &  1.5  \\
non-H bond lengths (MGNHBL11) &  7.6  &  9.2  &&  {\bf 7.2}  &  $^*$7.3        &  $^*$7.3         &&   9.8     &  8.8  &  8.7     &  10.7  &  10.6  \\
open-shell molecules (BL9)    & 15.0  & 14.7  && 11.9         & $^*${\bf 11.6}   & $^*${\bf 11.6}  && 12.7     & 12.2  & 12.3     & 12.5    &	12.5 \\
vibrations (F38)              & 58.4  &  55.0  &&  {\bf 37.1}  &  $^*$44.2        &  $^*$44.2        &&  53.3  &  52.0  &  52.1    &  58.3  &  58.3  \\
RMAE                         &  2.00  &  1.88  &&  0.91        &  $^*${\bf 0.77}  &  $^*${\bf 0.77}  &&  1.00  &  0.86  &  0.82    &  0.96  &  0.96  \\
WMAE(\%)                     &  +135  &  +119  &&  -7        &  $^*${\bf -28}    & $^*${\bf  -28}   &&  0      &  -17   &  -23      &  -7  &  -7  \\
\multicolumn{13}{c}{Transition metal complexes (kcal/mol [for AUnAE kcal/(mol$\cdot$atoms)] and m\AA)} \\
atomization energy (TM10AE)  &  13.0  &  11.1  &&  13.4  &  12.1         &  $^*${\bf 11.0}   &&  14.4    &  15.8  &  14.0  &  14.1  &  11.4  \\
reaction energies (TMRE)       &  3.7  &  3.1  &&  10.6   &  9.6          &  $^*${\bf 8.4}   &&  9.0     &  9.5  &  9.5  &  11.5  &  9.2  \\
gold clusters atomiz. (AUnAE)  &  0.60  &  1.8  &&  5.9   &  $^*${\bf 3.8}  &  $^*${\bf 3.8}   &&  4.0     &  5.1  &  5.0  &  4.3  &  4.3  \\
bond lengths (TMBL)           &  13.5  &  13.0  &&  18.5  &  $^*${\bf 18.3}  &  $^*${\bf 18.3}  &&  21.2  &  20.6  &  20.5  &  21.8  &  21.8  \\
gold clusters bonds (AuBL6)  &  56.5  &  58.7  &&  94.2  &  36.6           &  36.5            &&  41.9   &  56.0  &  58.9  &   32.5  &  $^*${\bf 32.3}  \\
RMAE                        &  0.69  &  0.72  &&  1.34  &  0.92          &  $^*${\bf 0.88}    &&  1.00   &  1.15  &  1.13  &  1.03  &  0.94  \\
WMAE(\%)                   &  -30  &  -28    &&  +29    & -8               & $^*${\bf -12}    &&  {\bf 0}  &  +13  &  +11  &  +3  & -6  \\
\multicolumn{13}{c}{Non-covalent interactions (kcal/mol)}\\
hydrogen bonds (HB6)       &  0.38  &  0.32  &&  0.55      &  $^*${\bf 0.36}  &  $^*${\bf 0.36}  &&  0.52  & {\bf 0.36}  &  0.45        &  0.38  &  0.38 \\
dipole-dipole (DI6)        &  0.38  &  0.39  &&  0.88       &  $^*$0.38    &  $^*$0.38         && 0.36     &  0.38       &  {\bf 0.32}  &  $^*$0.38  &  $^*$0.38  \\
dihydrogen bonds (DHB23)   &  0.98  &  0.76  && {\bf 0.34}  &  $^*$0.66    &  $^*$0.66        &&  0.75     &  0.56     &  0.79           &  $^*$0.66  &  $^*$0.66  \\
charge-transfer (CT7)      &  2.7  &  2.4  &&  {\bf 0.79}  &  1.3       &  1.3           &&  1.3      &  1.0      &  1.2       &  $^*$1.2  &  $^*$1.2  \\
$\pi$-$\pi$ stacking (pp5)  &  2.1  &  2.2  &&  3.2      &  $^*$2.2    &  $^*$2.2       &&  2.1      &  2.2      & {\bf 2.0}  & $^*$2.2  & $^*$2.2  \\
various non-covalent (S22)  &  2.3  &  2.7  &&  3.5      &  $^*$2.4    &  $^*$2.4       &&  {\bf 2.2}  &  2.4       &  {\bf 2.2}  &  $^*$2.4  & $^*$2.4  \\
RMAE                        &  1.21  &  1.13  &&  1.27  &  0.97        &  0.97          &&  1.00     &  {\bf 0.91}  &  0.95      &  $^*$0.95  &  $^*$0.95  \\
WMAE(\%)                    &  +23  &  +15  &&  +22      &  -5         & -5            &&  0         & $^*${\bf -12}  &  -6       & $^*$-7  & $^*$-7  \\
\multicolumn{13}{c}{Other (kcal/mol; Debye/10 for DM25)} \\
isomerization (ISOL6)    &  2.2   &  2.4  &&  2.7      &  $^*${\bf 1.4}  & $^*${\bf 1.4}  &&  1.5     & {\bf 1.4}  & {\bf 1.4}  & $^*${\bf 1.4}  & $^*${\bf 1.4}  \\
difficult cases (DC9.12)  &  40.8  &  29.4  &&  25.0    &  19.7         &  19.6      &&  {\bf 17.7}  &  22.9     &  20.2    &  18.2            &  $^*$17.9  \\
small interfaces (SI12)  &  3.7  &  5.9    &&  11.0    &  $^*${\bf 7.0}  &  7.1      &&  7.2        &  8.9      &  9.3     &  7.8           & 7.9  \\
dipole moments (DM25)   &  3.6   &  3.6    &&  2.9     &  2.7          &  2.7       &&  2.7       &  2.7       &  2.7      &  $^*${\bf 2.6}  & $^*${\bf 2.6}  \\
atomic energies (AE17)  &  51.6  &  22.2  && {\bf 7.6}  &  16.9       &  16.6       &&  42.2      &  10.2      &  54.5     &  15.7           &  $^*$15.3  \\
RMAE                    &  1.36  &  1.19  &&  1.20     &  0.88       &  0.88        &&  1.00      &  0.94      &  1.13     &  0.88          &  $^*${\bf 0.87}  \\
WMAE(\%)                & +38    &  +9     &&  +2     &  -22          & -23         &&  0        & -21         &  +17       & $^*${\bf -24}  & $^*${\bf -24}  \\
  \\
Global RMAE          &  1.41  &  1.27     &&  1.12     &  0.92      &  $^*${\bf 0.90}  &&  1.00  &  1.02      &  1.03     &  0.99  &  0.94  \\
Global WMAE(\%)      &  +44  &  +28      &&  +5        &  -11        & $^*${\bf -13}   &&  0     & -4         &  +2       & -5     &  -9    \\
%
\end{tabular}
\end{small}
\end{ruledtabular}
\end{center}
\end{table*}
Table \ref{tab1} reports the mean absolute error (MAE) for each test as obtained
from the different functionals. 
We see that the best overall performance is given by the 
hybrid functional defined by Eq. (\ref{e41}) with $n=5$ (named hAPBE hereafter),
which yields a global RMAE of 0.90 and a global WMAE of -13\%. 
Notably it is also the best non-fitted hybrid functional for
three out of five groups of tests that we considered
(thermochemistry, organic molecules' geometry, and transition metals) 
and it is very close to the best results for the remaining
two groups. Moreover, hAPBE provides results more
accurate than the average of the considered hybrids in 23
over 28 cases.
For thermochemistry the hAPBE functional yields performance very close to  
the B3LYP functional (RMAE is the same; WMAE is only 2\% better in B3LYP)
which was fitted on some of these properties.
However B3LYP shows, several limitations 
for other problems, especially those related to transition metals 
and clusters, being below the average performance in most tests
not connected to thermochemistry.
These limitations may possibly trace back to the
fitting parameters as well to some limitations of the 
Lee-Yang-Parr correlation \cite{lyp} in the slowly-varying density limit
\cite{paier07}.

We also note that the hAPBE functional is also superior to some
non-empirical meta-GGAs, such as TPPS \cite{tpss} and BLOC
\cite{loc,bloc,blochole} which have total WMAEs +16 and +13
respectively,  and simple hybrid meta-GGAs 
(TPPSh \cite{tpssh} has WMAE +3). 
It is still not as accurate as the most sophisticated hybrid meta-GGAs.
Hence, for example the M06 functional \cite{m06} performs
on the test set used in this work with a WMAE of -17\%. However, we need to stress
that the M06 functional contains several tens of empirical parameters.

\begin{figure*}
\includegraphics[width=\textwidth]{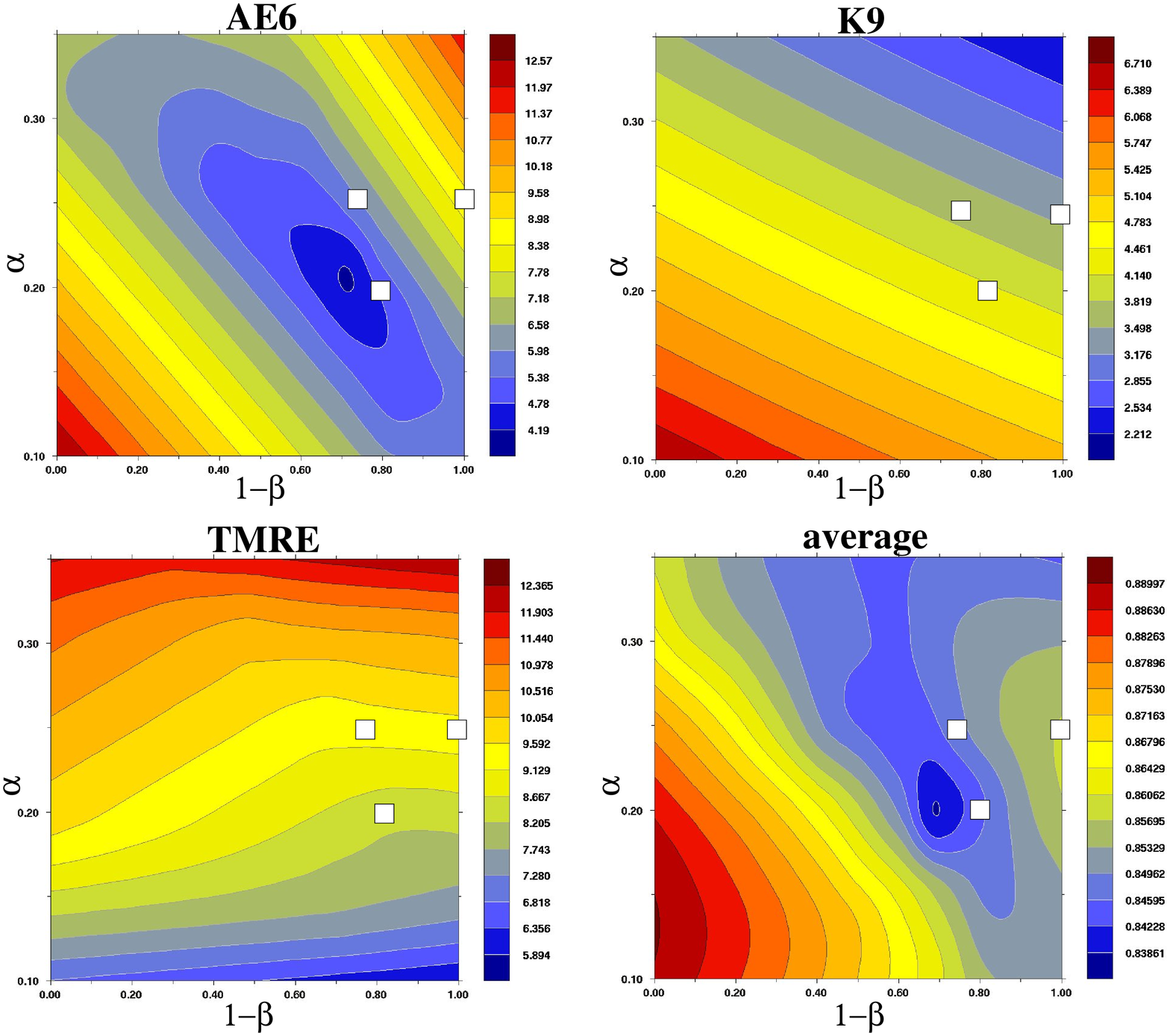}
\caption{\label{fig1} Mean absolute errors (kcal/mol) for several energy tests and different values of the parameters in the hybrid 
functional of Eq. (\ref{e6}). The scaled average (see text for the definition) of the various tests is also reported (right bottom 
panel). In each panel the white boxes denote the positions of the functionals
defined by Eq. (\ref{e6}) with $n=4$ and $n=5$ as well as APBE0.}
\end{figure*}

To understand better the results of Tab. \ref{tab1} two main trends are worth of
an analysis.
The first one concerns the effect of the inclusion of different amounts
of Hartree-Fock exchange.
Information on such a trend can be obtained comparing
APBE with APBE0, PBE with PBE0, and in general the hybrids
with $n=5$ and $n=4$.
The analysis of the data in Tab. \ref{tab1} shows that
in fact the inclusion of a fraction of Hartree-Fock exchange
can benefit the overall performance, but a larger
amount of Hartree-Fock exchange does not necessarily correspond
to an improvement of the results. More in detail we
see that non-covalent complexes are almost insensitive
to the inclusion of Hartree-Fock exchange (with the partial
exception of charge-transfer complexes) whereas organic
molecules and transition-metal complexes have different and quite
erratic behaviors. Hence, for organic molecules 
Hartree-Fock exchange provides an improvement for barrier heights and
most reaction energies.
On the other hand,
the inclusion of Hartree-Fock exchange yields a clear worsening of
transition-metal complexes energies, but a moderate improvement for the
bond lengths.
Thus, a proper balance, taking into account all these effects, appears
hard to find. Nevertheless, it seems that a best overall
performance may be obtained for a moderate fraction of
Hartree-Fock exchange. This result is in agreement with the finding
of Ref. \citenum{hpbeint}.
On the other hand, even within functionals having the same amount of
Hartree-Fock exchange, significant differences are found.
In particular, we remark that from the present assessment 
the hybrids proposed in this work appear to perform better 
than the popular hybrids B3LYP and PBE0.

The second trend to observe concerns the different possible choices
for the correlation part of the functional. In particular,
it is interesting to inspect the APBE0, PBEmol$\beta$0, and the functional
defined by Eq. (\ref{e4}) with $n=4$ (the spin-dependent
correction for PBEloc is discussed below),
to understand the role of the LDA linear response constraint 
and of the PBEloc correlation.
In agreement with Ref. \citenum{delcampo12} we find that
the imposition of the LDA linear response on the semilocal 
part of the functional gives in general an improvement
of the performance, especially for atomization and 
reaction energies of organic molecules. However, 
in the case of PBEmol$\beta$0 this seems to occur mostly 
thanks to an error cancellation effect, as shown by the
fact that atomic energies are much worse for PBEmol$\beta$0
than for APBE0. On the other hand, the functional proposed in this work 
shows a similar (but systematically better) 
performance as that of PBEmol$\beta$0, with apparently 
a smaller error cancellation.

We remark once more that neither APBE0 nor PBEmol$\beta$0 recover the full
LDA linear response, as instead do the functionals of the present work. 
In fact, the exchange-correlation second-order gradient coefficients are:
$\mu_{xc}^{PBE0}=-0.024$, $\mu_{xc}^{APBE0}=-0.034$, $\mu_{xc}^{PBEmol\beta 0}=0.031$,
while only for the functionals of Eqs. (\ref{e4}) and (9),  $\mu_{xc}=0$. 
Because in Table II the latter functionals give the best 
global RMAE and global WRMAE, the imposition of the full 
LDA linear response for the full global hybrid seems to be important.
This correlation issue may appear, at first sight to have
only a minor impact on most calculations, because energy differences are often
considered. However, it may be relevant for those cases
where heterogeneous systems are involved, such as for the atomization of metal clusters
or the description of metal-organic interfaces,
because in these cases the correlation effects of the different parts are
likely to be very different and will not cancel properly.
This observation is supported by the trends registered in Tab. \ref{tab1}
for the AUnAE and SI12 tests, although the small dimensions of the
systems considered in those tests prevent a good analysis of
this effect.

Finally, we consider a comparison between hybrid functionals
including the simple PBEloc correlation and those
implementing the spin-dependent correction via
zvPBEloc.
The data of Table \ref{tab1} show the latter functionals
display a non-negligible improvement with respect
to the former ones: the improvement in WMAE is 4\%  (2\%) for $n=4$  ($n=5$).
This fact can be traced back to the
higher compatibility with Hartree-Fock 
exchange of zvPBEloc with respect
to PBEloc. In fact, the more significant
improvement is registered for $n=4$, while smaller
advantages are observed for $n=5$.

Also interesting to note is the fact
that the inclusion of the spin-dependent
correction brings in all cases either an improvement
of the results or leaves them practically unchanged
(a very small worsening is observed only in few cases).
We recall, in addition, that for spin-compensated cases
(e.g. all the non-covalent tests) the
spin-correction has no effect, by construction.

\subsection{Analysis of hybrid parameters}\label{sec:par}

To gain more insight into the performance of the hybrid functionals
we consider in this section the general expression in 
Eq. (\ref{Final1}), with DFA1=APBE and DFA2=PBEloc, i.e.
\begin{equation}\label{e6}
E_{xc} = \alpha E_x^{HF} + (1-\alpha)E_x^{APBE} + \beta E_c^{PBEloc} + (1-\beta)E_c^{APBE}\ ,
\end{equation}
and we preform a scan of the values of the two parameters
$\alpha$ and $\beta$ to investigate the dependence of
the results on the fraction of Hartree-Fock exchange and the
correlation contribution.
To this end we consider a minimal set of tests composed of the
energy tests AE6 (atomization energies of small molecules),
K9 (barrier heights and reaction energies of organic molecules),
and TMRE (reaction energies of transition-metal complexes). 
These tests are in fact representative of the most important systems and properties 
as well as of the different trends observable in Table \ref{tab1}.
To evaluate a global performance of the energy tests, which display
rather different MAEs, we consider
an average of the three tests after rescaling the result of each 
test as MAE$\rightarrow$MAE/(1+MAE). Note that the scaling has 
the two-fold purpose of making all the quantities comparable, so
that a simple average is meaningful, and to enhance the differences
between small values (which are the most interesting here).

The results of the scan for the energy tests are reported in Fig. \ref{fig1}.
The plots show that indeed, as noticed when discussing Tab. \ref{tab1},
the different tests display different behaviors with respect to $\alpha$ and $\beta$.
In particular, 
the barrier heights within the K9 test
require a large fraction of Hartree-Fock exchange
for an accurate description,
while the opposite occurs in transition metals.
On the other hand, atomization energies of organic molecules show an even more
complex trend, requiring a delicate balance between exact exchange and correlation.
Thus, overall the average performance has a complicated dependence on the
two parameters. Nevertheless, it is possible to identify a clear minimum 
approximately corresponding to $\alpha=0.2$ and $\beta=0.3$, which is
close to the definition of the functional with $n=5$.
Moreover, reasonably small average errors can be obtained for 
the family of parameters respecting the relation $\beta\approx\alpha+0.1$,
which is close to the condition used in Eq. (\ref{e4}).
Such a relation also shows that, as discussed in Section \ref{sec:con}, an accurate correlation functional to be used in hybrid functional 
should depend on the HF exchange contribution.

\begin{figure}
\includegraphics[width=\columnwidth]{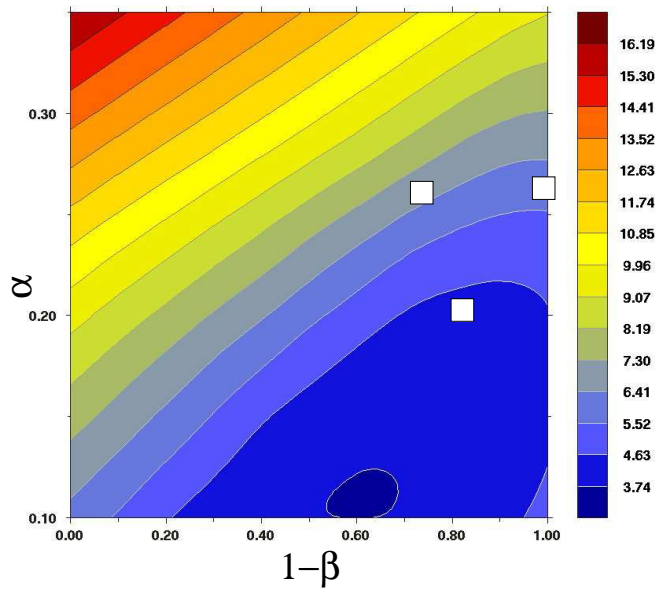}
\caption{\label{fig2} Mean absolute errors (m\AA) for geometry tests as obtained for different values of the parameters in the 
hybrid functional of Eq. (\ref{e6}). The white boxes denote the positions of the functionals defined by Eq. (\ref{e6}) with $n=4$ and $n=5$ as
well as APBE0.}
\end{figure}

These results indicate the robustness of our construction
based on the ansatz of Eq. (\ref{Final1}) and the three
imposed constraints. In particular, it is highlighted the
importance of the satisfaction of the LDA linear response
for the full XC functional, which is the constraint
determining the condition $m=n$ in the functional
of Eq. (\ref{e4}).
   
A somehow different situation is found concerning the geometry 
tests (MGHBL9 and MGNHBL11),
which are reported in Fig. \ref{fig2}. In this case 
in fact a large and shallow minimum
is observed for moderate values of the parameter $\alpha$
and $\beta$ values ranging from 0.2 to 1.
This indicates that reasonably small errors
can be achieved by hybrid functionals including a sufficiently
moderate fraction of Hartree-Fock exchange.
In fact, for $n=5$, the result of 4.4 m\AA{} is reasonably close to the
global minimum of 3.7 m\AA. 

Note also that the lines $\beta=1$ in both Figs. \ref{fig1} and \ref{fig2}, which correspond
to the hybrids which use the full PBEloc correlation, show a modest performance
for all the tests. Such hybrids not only violates the LDA linear response, but also
behave in the tail of the density as a pure exchange functional (because of the PBEloc
construction). We recall that the correlation energy density plays an important 
role for asymptotic properties \cite{PC1,PC2}. 
Nevertheless, remarkably both Figs. \ref{fig1} and \ref{fig2} show that the best results
are found for $\beta\approx 0.2$, which reveals an important significance of the PBEloc correlation
as a tool for the construction of hybrid functionals. 
Thus, further development of semilocal correlation functionals
more compatible with HF exchange, may further improve the quality of global hybrids.

\subsection{Semiempirical dispersion correction}

\begin{table}
\begin{center}
\caption{\label{tab2} Numerical values of the parameters used in the semiempirical D3 dispersion correction of the different 
functionals.}
\begin{ruledtabular}
\begin{tabular}{lll}
Functional & $s_{r,6}$ & $s_{8}$ \\
\hline
APBE   & 1.242 & 0.930 \\
APBE0  & 1.259 & 0.921 \\
this work $n=4$ & 1.266 & 0.915 \\
this work $n=5$ & 1.258 & 0.923 \\ 
\end{tabular}
\end{ruledtabular}
\end{center}
\end{table}

\begin{table*}
\begin{center}
\caption{\label{tab_d3} Mean absolute errors (MAEs) on different tests for the dispersion corrected (D3) hybrid functionals.}
\begin{ruledtabular}
\begin{tabular}{lrrrrrr}
 & \multicolumn{1}{c}{GGA} & \multicolumn{2}{c}{Hybrid $n=5$} & \multicolumn{3}{c}{Hybrid $n=4$} \\
\cline{2-2}\cline{3-4}\cline{5-7}
Test  &  APBE-D3 & Eq. (\ref{e4})-D3  & Eq. (\ref{e41})-D3  &  APBE0-D3  & Eq. (\ref{e4})-D3  & Eq. (\ref{e41})-D3 \\
hydrogen bonds (HB6)      &  0.77   &  1.0      &  1.0   &  1.0   &  1.1   &  1.1  \\
dipole-dipole (DI6)       &  0.91  &  0.73     &  0.73  &  0.69   &  0.69 &  0.69   \\
dihydrogen bonds (DHB23)  &  1.3  &  1.2    &  1.2     &  1.0    &  1.1 &  1.1   \\
charge-transfer (CT7)     &  2.9   &  1.8     &  1.8    &  1.5   &  1.6  &  1.6   \\
$\pi$-$\pi$ stacking (pp5)  &  0.2  &  0.18  &  0.18   &  0.20   &  0.21 &  0.21   \\
various non-covalent (S22)  &  0.5 &  0.59  &  0.59     &  0.57 &  0.62 &  0.62 \\
isomerization (ISOL6)  &  2.2  &  1.0  &  1.0  &  1.0  &  1.0 &   1.0   \\
small interfaces (SI12)  &  4.5   &  5.8  &  5.7  &  7.6  &  6.6  &  6.5 \\
\end{tabular}
\end{ruledtabular}
\end{center}
\end{table*}
To complete the construction of the hybrid functionals
we considered to complement them with a semiempirical
dispersion correction \cite{d3}. In fact, the dispersion interaction 
cannot be described neither by the Hartree-Fock exchange nor by any
semilocal correlation.
We implemented the dispersion correction through the DFT-D3 
semiempirical model \cite{d3}, fixing the two free parameters of 
the model by fitting to the MAE of the S22 test. 
The resulting parameters for different functionals are
listed in Table \ref{tab2}.
Note that because the S22 test only includes systems without any spin polarization
the parameters of the D3 correction are the same for functionals defined
by Eqs. (\ref{e4}) and (\ref{e41}).  

The D3 dispersion correction is found to be well compatible with all
the hybrid functionals, bringing a good improvement for dispersion
dominated cases (i.e. S22 and pp5, where van der Waals interactions 
dominate or rather large systems are considered; see Tab. \ref{tab_d3}).
Interestingly the
semiempirical dispersion correction also reduces the MAE for ISOL6 and 
metal-organic interfaces (SI12).
On the other hand, for the first four test in Tab. \ref{tab_d3} where dispersion is not dominating, 
the D3 correction tends to  slightly worsen the results for APBE and its hybrid variants.  
For all other tests (not reported Tab. \ref{tab_d3}) where dispersion does not play
any relevant role, results are correctly left almost unchanged (data non reported).

\section{Conclusions}
We have developed the global hybrids of the APBE 
exchange-correlation GGA functional \cite{apbe}.
The construction of these hybrids is based on the use 
of the PBEloc or zvPBEloc correlation functionals,
to accompany the fraction of exact exchange, and on 
the recovery of the accurate LDA linear response. 
By a broad energetical and structural testing, 
including thermochemistry, organic geometry, atomization energies, 
reaction energies, and bond lengths of transition metal complexes, 
non-covalent interactions, isomerization, gold-molecules hybrid-interfaces, 
and dipole moments of organic molecules, we have shown that 
the best total performance is obtained by considering the hybrid
using the zvPBEloc correlation (i.e Eq. (\ref{e41})) and
$n=5$. We name this functional hAPBE. The hAPBE hybrid functional, 
shows an almost systematic improvement over the popular PBE0 and B3LYP hybrids, 
as well as over the recently proposed PBEmol$\beta$0 functional.
In fact, the hAPBE hybrid functional has a good accuracy 
for all the tests, showing a broad applicability, 
in contrast to the B3LYP functional which is 
modest for the transition metal complexes.
Use of a semiempirical dispersion correction can bring further
accuracy for problems where the dispersion
interaction is especially relevant.
Furthermore, the analysis of the hybrid parameters, 
summarized in Figs. \ref{fig1} and \ref{fig2}, has shown that 
the LDA linear response is a powerful constraint, and that the PBEloc correlation
plays a significant role in the functional performance. 
Thus the hAPBE functional can be regarded as an interesting tool
for quantum chemistry applications. Whereas,
the construction presented in this work, using a different correlation
contribution to accompany the Hartree-Fock exchange and fulfilling
the LDA linear response, appears to be an important
strategy to develop more accurate hybrid functionals 
in the future.

In view of further improvements different 
semilocal correlation functionals with good compatibility with the
Hartree-Fock exchange can be considered (e.g. GAPloc \cite{gaploc} or revTCA {\cite{RevTCA}). 
Alternatively, the replacement of GGAs with meta-GGAs can be considered in Eq. (\ref{PEBmodg2}).
We recall in fact that important recent work at the meta-GGA level has appeared.
Hence, several accurate non-empirical (semi-empirical) functionals have been proposed, 
such as revTPSS \cite{revtpss}, regTPSS \cite{regtpss}, MGGA-MS \cite{mggams}, VT$\{8,4\}$ \cite{VT08},
and BLOC \cite{loc,bloc,blochole}. All of them represent good candidates for the 
construction of an accurate meta-GGA hybrid, using the method proposed in the
present paper. 
In particular the BLOC functional uses a correlation
term (TPSSloc \cite{loc}) which has been derived from the PBEloc GGA.
Thus, this functional appears as the most natural choice for future studies.

\section{Appendix}
In this Appendix we analyze the coupling-constant-resolved XC potential energy formula in 
Eqs. (\ref{PEBmod}) and (\ref{PEBmodg2}) for atomic systems.
As a reference we consider the coupling-constant-resolved XC potential energy
of the  
interaction-strength interpolation (ISI) model  ($W_{xc,\lambda}^{ISI}$), a high-level method constructed 
from exact conditions \cite{ISI}.

The differences between the various approximate $W_{xc,\lambda}$ and the reference,
as functions of $\lambda$, for several neutral atoms, are reported in \ref{fig_wxc}.
In addition we consider also the Ne$^{8+}$ ion as an example for the high-density limit
case.
For all atoms we report also the exact behavior $W_{xc,\lambda}\rightarrow
E_x+2\lambda E_c^{GL2}$, where $E_c^{GL2}$ is the 
second-order G\"orling-Levy perturbation theory correlation (GL2) \cite{gl2,ISI2}, 
for small values of $\lambda$. Note that for Ne$^{8+}$ this behavior becomes almost exact
over the whole range of $\lambda$ values.
\begin{figure*}
\includegraphics[width=0.8\textwidth]{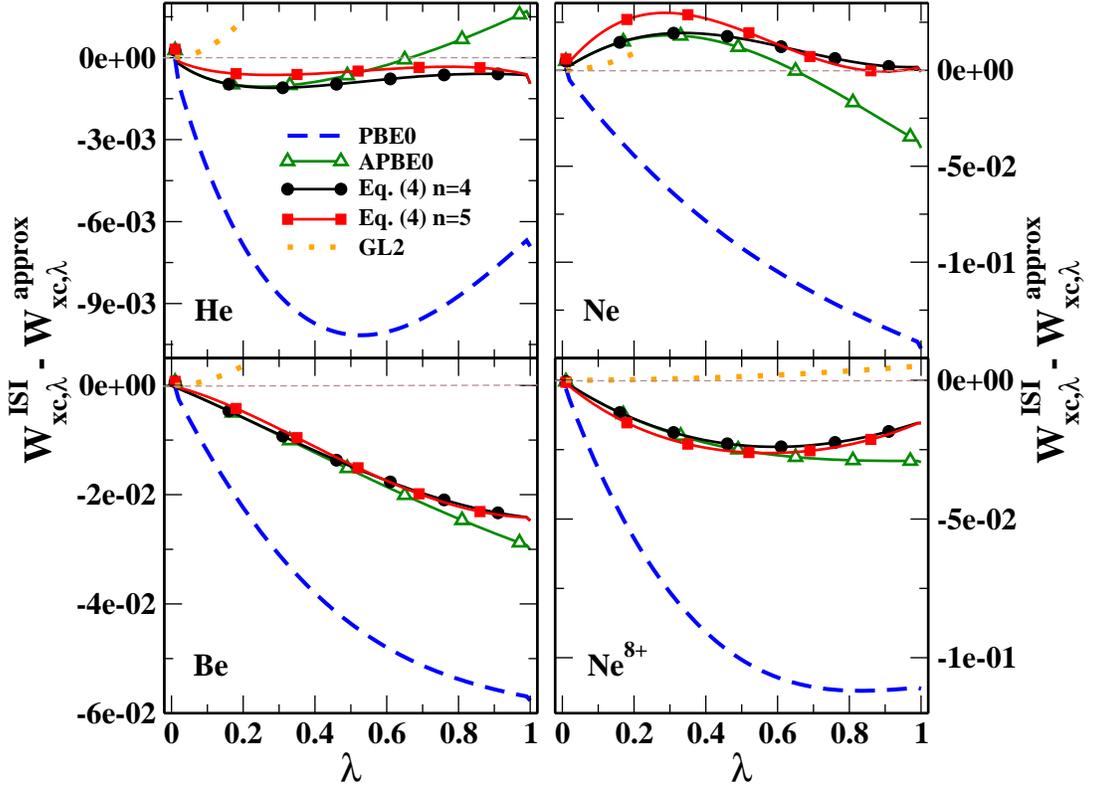}
\caption{\label{fig_wxc} Difference between several approximate coupling-constant-resolved XC 
potential energy formulas and the ISI model for the He, Be, Ne atoms and the Ne$^{8+}$ ion. 
For small values of $\lambda$ also the GL2 curve is shown as a reference for the initial slope.} 
\end{figure*}
The data in  \ref{fig_wxc} show that the PBE0 ansatz
is correct only at $\lambda=0$ and in the close proximity
of this point (by construction), whereas it shows a significant inaccuracy for
larger $\lambda$ values.
The results are much improved for all atoms considered when
the semilocal DFT approximation is changed from PBE to APBE
(APBE0 curve). In this case in fact a closer behavior
to the reference is obtained over the whole range of
$\lambda$ values. 
By construction, both $W_{xc,\lambda}^{PBE0}$ and $W_{xc,\lambda}^{APBE0}$ are exact at $\lambda=0$, 
while at $\lambda=1$, they recover the GGA behaviors, $W_{xc,\lambda}^{PBE0}\rightarrow 
W_{xc,\lambda}^{PBE}$ and $W_{xc,\lambda}^{APBE0}\rightarrow  W_{xc,\lambda}^{APBE}$ \cite{pbe0}. As 
shown in Fig. \ref{fig_wxc}, $W_{xc,\lambda}^{APBE}$ significantly improves over 
$W_{xc,\lambda}^{PBE}$ at $\lambda\rightarrow 1$, and consequently the APBE0 hybrid is more 
realistic. 

Then we consider the ansatz introduced in Eq. (\ref{PEBmodg2}).
The curves labelled ``$Eq. (4) n=4$'' and  ``$Eq. (4) n=5$'' in  \ref{fig_wxc} 
clearly show that the correlation correction has a non negligible effect only for $\lambda > 0.6$ and that, for all system considered, a 
significant improvement is obtained 
for $\lambda$ close to 1.
In particular 
an almost vanishing error
for  $\lambda=1$ is obtained for He and Ne.

\end{document}